\documentclass[aps,prb,reprint,showkeys,showpacs,amsfonts,amsmath,amssymb,superscriptaddress]{revtex4-1}

\usepackage[ansinew]{inputenc}
\usepackage{graphicx}

\usepackage{pgfplots}
\usepackage{pgfplotstable}
\usepgfplotslibrary{external} 
\tikzset{external/system call={latex \tikzexternalcheckshellescape  -interaction=batchmode -jobname "\image" "\texsource" && dvips -o "\image".eps "\image".dvi}}

\pgfplotsset{compat=1.5}

\begin{document}

\title{Rayleigh-B\' enard Instability in Graphene}

\author{O. Furtmaier}
\email{oliverfu@ethz.ch}
\affiliation{ETH Z\"{u}rich, Department of Physics, Wolfgang-Pauli-Strasse 27, HIT, 8093 Z\"{u}rich, Switzerland}
\author{M. Mendoza}
\email{mmendoza@ethz.ch}
\affiliation{ETH Z\"{u}rich, Computational Physics for Engineering Materials, Institute for Building Materials, Schafmattstrasse 6, HIF, 8093 Z\"{u}rich, Switzerland}
\author{I. Karlin}
\affiliation{ETH Z\"{urich}, Department of Mechanical and Process Engineering, Sonneggstrasse 3, ML K 20, CH-8092, Switzerland}
\author{S. Succi}
\affiliation{Istituto per le Applicazioni del Calcolo C.N.R, Via dei Taurini 19, 00185 Rome, Italy}
\author{H. J. Herrmann}
\affiliation{ETH Z\"{u}rich, Computational Physics for Engineering Materials, Institute for Building Materials, Schafmattstrasse 6, HIF, 8093 Z\"{u}rich, Switzerland}
\altaffiliation{Departamento de F\' isica, Universidade Federal do Cear\' a, Campus do Pici, 60455-760 Fortaleza, Cear\' a, Brazil}

\begin{abstract}
Motivated by the observation that electrons in graphene, in the hydrodynamic regime of transport, can be treated as a two-dimensional ultra-relativistic gas with 
very low shear viscosity, we examine the existence of the Rayleigh-B\'enard instability in a massless electron-hole plasma. 
Firstly, we perform a linear stability analysis, derive the leading contributions to the relativistic Rayleigh number, and calculate the critical value above 
which the instability develops. By replacing typical values for graphene, such as thermal conductivity, shear viscosity, temperature, and sample sizes, 
we find that the instability might be experimentally observed in the near future. Additionally, we have performed simulations for vanishing 
reduced chemical potential and compare the measured critical Rayleigh number with the theoretical prediction, finding good agreement. 
\end{abstract}


\maketitle

\section{Introduction}
Graphene, consisting of literally a single carbon monolayer, represents the first instance ever of a truly two-dimensional material. 
Due to the special symmetries of the honeycomb lattice, electrons in graphene are shown to behave like an effective Dirac fluid of massless, 
chiral quasi-particles, propagating at a Fermi speed of about $v_F \approx 1.1\times 10^6$m/s, cf. Refs. \cite{geim04,geim05}. 
Recent theoretical insights \cite{mueller08a,mueller08b,fritz08,mueller09} have opened up the possibility to describe the electrical current 
on graphene close to the Dirac point as a classical relativistic fluid when inelastic electron-electron and electron-hole scattering dominate 
over elastic electron-impurity and electron-phonon processes. Since electrons are about $300$ times slower than photons, their mutual interaction 
is proportionately enhanced, leading to an effective fine-structure constant $\alpha_{gr} = e^2/\hbar v_F \sim 1$.  As a result of such strong interactions, 
it has been recently proposed that this peculiar 2D graphene electron gas should be characterised by an exceptionally low viscosity/entropy ratio (near-perfect fluid), 
coming close to the famous AdS-CFT lower bound \cite{kovtun05} conjectured for quantum-chromodynamic fluids, such as quark-gluon plasmas.
This spawns the exciting prospect of observing electronic pre-turbulence in graphene samples, as first pointed out in Ref. \cite{mueller09} and 
confirmed by recent numerical simulations \cite{mendoza11}.\\
One of the most common hydrodynamic instabilities, is the Rayleigh-B\' enard, which has been studied experimentally since 1901 \cite{benard01}, 
and first described theoretically by \textit{Lord Rayleigh} in 1916 \cite{rayleigh16}. Here, one observes the creation of convection cells, 
which couple the heat and particle flow, where the driving mechanisms are the temperature gradient and the buoyancy force 
(typically caused by gravity \cite{chandra61}). This instability appears in many scientific and industrial applications \cite{succi93,ahlers12,moon14}, 
and is still subject of current research.\\
Since electrons and holes in graphene can be described by relativistic hydrodynamics, it opens the question 
whether the Rayleigh-B\'enard instability can be also observed. Although gravity in this case is negligible, 
we can use an external electric field to mimic its effects. In the classical case, the Rayleigh-B\' enard instability will always appear 
if the Rayleigh number \cite{chandra61}, defined by 
\begin{equation}
\mathrm{Ra}_\text{cl} = \frac{\alpha g \beta \rho c_V l^4}{\kappa \eta} \text{ ,}
\end{equation}
exceeds a critical value. Here, $\alpha$ is the thermal volume expansion coefficient, $g$ the gravitational acceleration, 
$\beta$ the uniform adverse temperature gradient, $c_V$ the specific heat at constant volume, $\rho$ the mass density, $l$ the length of the system parallel 
to the gravitational force, $\kappa$ the thermal conductivity, and $\eta$ the shear viscosity. From this expression, we can notice that for low values 
of shear viscosity and thermal conductivity, we obtain a large Rayleigh number, and consequently, they contribute to the appearance of the instability. 
Graphene's electronic fluid has a very low shear viscosity but a high thermal conductivity \cite{balandin08}, therefore, it is not straightforward to 
judge the existence of the instability, under actual experimental conditions. In order to answer this question, we first introduce a linear stability analysis 
in the context of relativistic hydrodynamics, which has not been formulated before, to the best of our knowledge, and calculate the critical 
relativistic Rayleigh number. We also perform numerical simulations of electronic flow in graphene in order to observe the convection rolls, and to 
validate our linear stability analysis. By studying the parameter regime, in which the instability can be observed in graphene, we have found that under 
actual experimental conditions, it is unlikely to observe the instability, but since the construction of larger samples of graphene is a target in current research, 
it might be observed in the near future.\\
This work is organised as follows. In Section \ref{Hydro} we write the macroscopic equations from the relativistic Boltzmann transport theory for graphene in 
the ``hydrodynamic regime" \cite{fritz08}. These  partial differential equations build the basis for the linear stability analysis, which will be derived in 
Section \ref{theory}. Thereby, we define the dimensionless relativistic Rayleigh number, determine its critical value, which marks the onset of the instability, 
and investigate the experimental realizability of the phenomenon. The Einstein summation convention and the signature $(+,-,-)$ for the Minkowski metric 
$g^{\alpha\beta}$ are used. Greek indices run from 0 to 2 whereas Latin indices can only take the values 1 and 2. The other constants which are used throughout 
the article are $k_B$ for the Boltzmann constant, $\hbar\equiv h/2\pi$ for the reduced Planck constant, and $e$ for the absolute value of the elementary electric 
charge. In Section \ref{Sim}, we perform simulations of the electron flow for vanishing reduced chemical potential. For this purpose, we first extend the 
relativistic lattice Boltzmann developed in Ref. \cite{oettinger13}, analyse the functional form of the shear viscosity and thermal conductivity, 
and validate our implementation of the external force. Finally, the critical relativistic Rayleigh number is measured and compared to the theoretical prediction. 
As a general introduction to the lattice Boltzmann method and its extension to relativistic dynamics and quantum statistics we refer the reader to Refs. 
\cite{succi92}, \cite{mendoza10} and \cite{uehling33}.

\section{Hydrodynamic regime overview}
\label{Hydro}
In the hydrodynamic regime \cite{mueller09,fritz08}, it is known that carriers in graphene behave as a two-dimensional relativistic fermionic plasma 
whose quasi-particles are charged, nearly massless, and moving at the Fermi speed $c\equiv v_F$. Henceforth, 
we will denote the Fermi speed by $c$. In order to describe the collision-dominated electron-transport in this material, we use the relativistic Boltzmann 
equation \cite{mueller09}
\begin{equation}
p^{\alpha}\frac{\partial f_{\sigma}}{\partial x^{\alpha}} + \sigma K^{\alpha}\frac{\partial f_{\sigma}}{\partial p^{\alpha}} = \mathcal{J}^{(\sigma)}_{\text{coll}}\label{RBE}\text{ ,}
\end{equation}
with $\sigma=\mp$ denoting electrons ($+$) and holes ($-$), $p^{\alpha}$ being the (pseudo) 3-momentum and 
$\sigma K^{\alpha} = \sigma F^{\alpha\beta}p_{\beta}$ being the electromagnetic 3-force acting on the electron/hole fluid component 
with distribution function $f_{\sigma}$ in phase-space $\left(\vec{x},\vec{p}\right)$ at time $t$ \cite{kremer02}. 
Here, $J^{(\sigma)}_{\text{coll}}$ denotes the collision operator, further explained in Ref. \cite{fritz08}. $F^{\alpha\beta}$ is the ``electromagnetic field strength tensor", 
defined as
\begin{equation*}
F^{\alpha\beta} = \begin{pmatrix}
0 & eE_1/c & eE_2/c\\
-eE_1/c & 0 & eB \\
-eE_2/c & -eB & 0 
\end{pmatrix}\text{ ,}
\end{equation*}
with $\vec{E}=(E_1,E_2,0)$ and $\vec{B}=(0,0,B)$ being the electric and magnetic fields, respectively. This formulation implies that the Coulomb interactions 
are invariant under Lorentz transformations defined with the speed of Fermi. Obviously, in fact they spread at the real speed of light, $c_L \gg c$, and therefore, 
they break the Lorentz invariance of our system. However, since we will deal with fluid flows at low speeds we can work in the laboratory frame, cf. Ref. \cite{mueller09}. 
The contravariant ``charge flow" $J^{\alpha}$ and energy-momentum tensor $T^{\alpha\beta}$ are defined as first and second moment of the distribution function:
\begin{align}
J^{\alpha}(\vec{x},t) &\equiv \langle p_{\sigma}^{\alpha} \rangle_{\sigma} \equiv \sum_{\sigma} c\int\!\frac{\mathrm{d}\vec{p}}{p_0}\hspace{1ex} p_{\sigma}^{\alpha}f_{\sigma}(t,\vec{x},\vec{p})\text{ ,}\label{def0}\\
T^{\alpha\beta} &\equiv \langle p_{\sigma}^{\alpha}p_{\sigma}^{\beta}\rangle_{\sigma} \equiv \sum_{\sigma} c\int\!\frac{\mathrm{d}\vec{p}}{p_0}\hspace{1ex} p_{\sigma}^{\alpha}p_{\sigma}^{\beta}f_{\sigma}(t,\vec{x},\vec{p})\text{ ,}\label{def1}
\end{align} 
with $p_{\sigma}^{\alpha} = \sigma p^{\alpha}$.
Using Eq.~\eqref{RBE}, the definitions \eqref{def0}, \eqref{def1}, and integrating by parts, the macroscopic conservation equations for charge, energy and momentum can be obtained,
\begin{align}
\partial_{\alpha}J^{\alpha} &= 0 \text{ ,}\label{PC}\\
\partial_{\beta}T^{\alpha\beta} &= F^{\alpha\nu}J_{\nu}\text{ .}\label{EMC}
\end{align}
Following an idea expressed in Ref.~\cite{landau87}, we employ the Eckart decomposition, cf. Ref.~\cite{kremer02}, with respect to the 3-velocity 
$U^{\alpha}\equiv \frac{\left(c,u_1,u_2\right)^{\alpha}}{\sqrt{1-|\vec{u}|^2/c^2}}$ of the ``charge flow" $J^{\alpha}$, cf. Ref.~\cite{wang86}, 
i.e. the current 3-vector and the energy-momentum tensor for the case of the viscous heat conducting fluid are given by 
\begin{align}
J^{\alpha} =\hspace{0.5ex}& n U^{\alpha} \text{ ,}\label{J0}\\
T^{\alpha\beta} =\hspace{0.5ex}& \epsilon\frac{U^{\alpha}U^{\beta}}{c^2} - \left(p+\omega\right)\Delta^{\alpha\beta}+p^{<\alpha\beta>}\notag\\
&+\frac{1}{c^2}\left(U^{\alpha}q^{\beta}+U^{\beta}q^{\alpha}\right)\text{,}\label{T00}
\end{align}
where $n(\vec{x},t)$ is the charge density divided by $e$, $\epsilon$ the internal energy density, $p$ the hydrostatic pressure, $\omega$ the dynamic pressure, $q^{\alpha}$ the contravariant heat flux and $p^{<\alpha\beta>}$ the pressure deviator in local equilibrium. Assuming that linear irreversible thermodynamics can be applied and using the procedure outlined in Ref. \cite{kremer02}, we find for the constitutive equations 
\begin{align}
p^{<\alpha\beta>} &= 2\eta\left[\frac{1}{2}\left(\Delta^{\alpha}_{\gamma}\Delta^{\beta}_{\delta}+\Delta^{\alpha}_{\delta}\Delta^{\beta}_{\gamma}\right)-\frac{1}{3}\Delta^{\alpha\beta}\Delta_{\gamma\delta}\right]\nabla^{\gamma}U^{\delta}\text{ ,}\notag\\
q^{\alpha} &= \kappa \left(\nabla^{\alpha}T-\frac{T}{c^2}DU^{\alpha}\right)\text{ ,}\label{hflux}\\
\omega &= -\nu\nabla_{\alpha}U^{\alpha}\text{ ,}\notag
\end{align}
where $\eta$ and $\nu$ stand for the shear and bulk viscosities, and $\kappa$ for the thermal conductivity. $\Delta^{\alpha\beta}\equiv g^{\alpha\beta}-\frac{U^{\alpha}U^{\beta}}{c^2}$ stands for the projector into the space perpendicular to $U^{\alpha}$, $D\equiv U^{\alpha}\partial_{\alpha}$ for the convective time-derivative, and $\nabla^{\alpha}\equiv\Delta^{\alpha\beta}\partial_{\beta}$ for the gradient operator in this decomposition.  It should be noted that the direct contribution of the electromagnetic fields to the heat flux vanishes due to the antisymmetry of the field strength tensor. However, they have an indirect contribution through the term $DU^{\alpha}$.
\subsection{Non-relativistic current flow}
Assuming $|\vec{u}|/c \ll 1$, we simplify the conservation equations (\ref{PC}) and (\ref{EMC}) to
\begin{align}
0 &=\partial_t n + \vec{\nabla}\cdot\left(n\vec{u}\right)\label{PC1}\text{ ,}\\
\varphi &=\partial_t \epsilon + \vec{\nabla}\cdot\left[(\epsilon+p)\vec{u}\right]+\vec{\nabla}\cdot\vec{q}+n\vec{u}\cdot e\vec{E}\label{EC1}
\text{ ,}\\
\vec{\psi} &=\frac{\partial_t}{c^2}\left[\left(\epsilon+p\right)\vec{u}+\vec{q}\right]+\vec{\nabla}p+n e\vec{E}\label{MC1}\text{ ,}
\end{align}
with $\varphi \equiv \partial_i \left[\left(\nu-\frac{2\eta}{3}\right)u_i\left(\vec{\nabla}\cdot\vec{u}\right)+\eta\vec{u}\cdot\left(\vec{\nabla}u_i + \partial_i \vec{u}\right)\right]$ being the change in energy due to dissipation and $\psi_i\equiv -\partial_i\left[\left(\frac{2\eta}{3}-\nu\right)\vec{\nabla}\cdot\vec{u}\right]+\partial_j\left[\eta\left(\partial_i u_j + \partial_j u_i\right)\right]$ the viscous term. To get those equations we have used the approximations from Ref. \cite{kremer02} for the expressions \eqref{J0} and \eqref{T00}, and recovered the Fourier law of heat conduction $\vec{q}=-\kappa\vec{\nabla}T$.
\section{Linear Stability Analysis}
\label{theory}
In analogy to the classical Rayleigh-B\' enard problem \cite{severin01}, we study a system with constant volume, confined to $(x_1,x_2)\in \mathbb{R}\times [-l/2,l/2]$ with $l$ being the length of our system. It has two thermal contacts at $x_2 = \pm l/2$ with temperatures $T_{\pm}$, respectively, and feels a constant homogeneous external electric field $\vec{E}_{\text{ext}}=E_{\text{ext}}\hat{e}_2$. We consider our system thermally isolated elsewhere, which can be achieved, for example, by a freely suspended sheet \cite{seol10}, or using supports with a much smaller thermal conductivity. However, the contact with another material may require to consider a graphene-substrate interaction which can alter the dynamics. For the description of the electron-hole fluid we choose the velocity $\vec{u}$, the temperature $T$, the volume $V$ and the electro-chemical potential $\mu$ as state variables. Following the procedure outlined in Ref. \cite{chandra61} we make a linear stability analysis by perturbing the 
stationary, non-homogeneous, quiescent state characterized by ($\vec{u}=0$, $T=\bar{T}$, $\mu=\bar{\mu}$). This state is chosen to simplify the theoretical analysis. By symmetry, the variables $\bar{T}$ and $\bar{\mu}$ only depend on $x_2/l$. They can be calculated by solving the ordinary differential equations
\begin{align}
\bar{T}'\kappa'(\bar{T},\bar{\mu}) + \bar{T}'' \kappa(\bar{T},\bar{\mu})&=0\text{ ,}\\
p'(\bar{T},\bar{\mu}) -l E_2(\bar{T},\bar{\mu})e n(\bar{T},\bar{\mu}) &= 0\text{ ,}\label{sol2}
\end{align}
which follow from Eqs. \eqref{EC1} and \eqref{MC1} under the stated assumptions. It should be noted that the average electric field 
$\bar{E}_2=E_2(\bar{T},\bar{\mu})$ not only contains external information, as the fluid components also experience an internal electric field due 
to intrinsic distributions of charge.

\subsection{Perturbation equations}
Writing for each fluid quantity in Eqs. \eqref{PC1} to \eqref{MC1}, $X=\bar{X}+\tilde{X}$ with $X \in \lbrace T,\epsilon,p,n,u_i,E_i,\kappa,\eta,\nu \rbrace$ and ignoring non-linear terms in the perturbations $\tilde{X}$ we find for the perturbation equations
\begin{align}
\partial_t \tilde{n}+\bar{n}' u_2 + \bar{n}\partial_i \tilde{u}_i &= 0\text{ ,}\label{CE1}\\
\partial_t \tilde{\epsilon}+\bar{\epsilon}'\tilde{u}_2+c^2\bar{\rho}\partial_i \tilde{u}_i+\partial_i \tilde{q}_i&= 0\text{ ,}\label{CE2}\\
\bar{\rho}\partial_t \tilde{u}_i+\frac{\partial_t \tilde{q}_i}{c^2}-
\partial_j\left[\bar{\eta}\left(\partial_i \tilde{u}_j + \partial_j \tilde{u}_i\right)\right]+\partial_i \tilde{p}+\tilde{f}_i&= \notag\\
\partial_i \left[\left(\frac{2\bar{\eta}}{3}-\bar{\nu}\right)\partial_j \tilde{u}_j\right]&\hspace{0.5ex}\text{  ,}\label{CE3}
\end{align}
where we have defined $\bar{\rho}\equiv \frac{\bar{\epsilon}+\bar{p}}{c^2}$ as mass density,
\begin{align*}
\tilde{q}_i &\equiv \bar{\kappa}\partial_i \tilde{T}+\tilde{\kappa}\bar{T}' \delta_{i2}\text{ , and}\\
\tilde{f}_i &\equiv \tilde{n}e\bar{E}_2\delta_{i2}+\bar{n}e\tilde{E}_i \text{ ,}
\end{align*}
for the perturbations of the heat flux and force density, respectively. In addition, we assume that the thermodynamics of the plasma is very similar to a photon gas. Hence, we know that its internal energy is mainly determined by the state variables of temperature $T$ and volume $V$, i.e. $\tilde{\epsilon}\approx \bar{c}_{V} \tilde{T}$ with $c_{V}$ being the volume-specific heat capacity at constant volume. Employing the Oberbeck-Boussinesq approximation~\cite{chandra61}, for which the electric field takes the role of the gravitational field, the perturbation equations in linear approximation become 
\begin{align}
\partial_i\tilde{u}_i =\hspace{0.5ex}& 0 \text{ ,}\\
\partial_t \tilde{T} =\hspace{0.5ex}& -\beta_R\tilde{u}_2+\frac{\kappa_R}{\left(c_{V}\right)_R}\Delta\tilde{T}\text{ ,}\\
\partial_t \tilde{u}_i =\hspace{0.5ex}& -\left(\alpha_{q}\right)_R a_R \tilde{T}\delta_{i2}-\frac{n_R}{\rho_R}e \tilde{E}_i+\frac{\eta_R}{\rho_R}\Delta\tilde{u}_i\notag\\
&-\frac{1}{\rho_R}\partial_i\left(\tilde{p}-\frac{\kappa_R}{c^2}\partial_t \tilde{T}\right)\text{ ,}\label{MC2}
\end{align}
with reference temperature $T_R=\bar{T}(x_2=0)$, gradient $\beta_R \equiv \bar{T}'(x_2=0)$, reference electro-chemical potential $\mu_R = \bar{\mu}(x_2=0)$, reference acceleration $a_R \equiv \frac{n_R e E_R}{\rho_R}$, electric field $E_R \equiv \bar{E}_2(T_R,\mu_R)$ and reference thermal charge expansion coefficient $\left(\alpha_q\right)_R \equiv \frac{1}{n_R}\left.\frac{\partial n}{\partial T}\right|_{T=T_R,\mu=\mu_R}$. The approximation will be reasonable if variations in these properties primarily stem from temperature fluctuations and $\frac{T_+-T_-}{T_+ +T_-} \ll 1$ holds. Higher order terms, which include non-Boussinesq effects, can also be considered by the method described in Ref.~\cite{herwig92}. In analogy to Ref.~\cite{chandra61}, we apply the curl operator twice on Eq.~\eqref{MC2} to get rid of the gradients and use the component parallel to the applied electric field
\begin{equation*}
\left(\frac{\eta_R}{\rho_R}\Delta-\partial_t\right)\left[\Delta \tilde{u}_2-\partial_2 \partial_i \tilde{u}_i\right] = \left(\alpha_q\right)_R a_R \partial_1^2 \tilde{T} -\frac{n_R}{\rho_R}e \partial_t \partial_1\tilde{B}\text{.}
\end{equation*}
From the Maxwell-Amp\` ere equation~\cite{jackson99}, we can deduce
\begin{equation*}
\partial_t \partial_1 \tilde{B} \approx -\mu_0 e n_R \partial_t\tilde{u}_2-\mu_0 \epsilon_0 \partial_t^2 \tilde{E}_2 \text{ ,}
\end{equation*}
with $\mu_0$ and $\epsilon_0$ being the vacuum permeability and permittivity, respectively. In this form, we realize that the contribution of this term to the overall dynamics is negligible, since $\frac{\mu_0 e^2 n^2_R}{\rho_R} \ll 1$ and $\frac{\mu_0\epsilon_0 e n_R}{\rho_R} \ll 1$. Consequently, the final and smaller set of simplified linear perturbation equations reads
\begin{align}
\partial_i \tilde{u}_i &=0\text{ ,}\\
\partial_t \tilde{T} &= \beta_R\tilde{u}_2+\frac{\kappa_R}{\left(c_V\right)_R}\Delta\tilde{T}\text{ ,}\\
\left(\alpha_q\right)_R a_R \partial_1^2 \tilde{T}&= \left(\frac{\eta_R}{\rho_R}\Delta-\partial_t\right)\Delta \tilde{u}_2\text{ ,}
\end{align}
and has the same form as in Ref.~\cite{chandra61}. Therefore, we define the Prandtl and relativistic Rayleigh number of this system as
\begin{align}
\mathrm{Pr} &\equiv \frac{\eta_R \left(c_V\right)_R}{\rho_R\kappa_R}\text{ ,}\\
\mathrm{Ra}_{\text{rel}} &\equiv \frac{\left(\alpha_q\right)_R a_R \beta_R l^4}{\frac{\kappa_R}{\left(c_V\right)_R}\frac{\eta_R}{\rho_R}} = \frac{\left(\alpha_q c_V \right)_R n_R e E_R\beta_R l^4}{\kappa_R \eta_R}\text{ ,}\label{RelRay}
\end{align}
by following the method of Ref.~\cite{severin01}. In analogy with the classical case, we expect a positive Rayleigh number. 
Therefore, we find that the possibility for the occurrence of the instability depends on the sign of the product of the temperature gradient, 
the external electric field and the thermal charge expansion coefficient, as all other parameters are non-negative. 

\subsection{Analysis into normal modes}

Since the equations in the previous subsection have the same form as in the classical case, except for different coefficients, 
we can draw on the work in Ref.~\cite{chandra61} concerning the analysis into normal modes and the calculation of the critical Rayleigh number. 
Due to the geometry of the problem we can expand any perturbation into the complete set of plane waves and hence write
\begin{align}
\tilde{T}(x_1,x_2,t) &= \int\limits_{\mathbb{R}}\!\mathrm{d}k \hspace{1ex} \theta_k (x_2) e^{s_k t + \imath k x_1}\text{ ,}\\
\tilde{u}_i (x_1,x_2,t) &= \int\limits_{\mathbb{R}}\!\mathrm{d}k \hspace{1ex} u_{k,i}(x_2) e^{s_k t + \imath k x_1}\text{ ,}
\end{align}
with $s_k \in \mathbb{C}$. One can show that the imaginary part of $s_k$ needs to vanish if one demands non-conducting or transversal-conducting boundaries at $x_2=\pm l/2$. 
These conditions imply for the charge velocity perturbations under consideration of $\partial_i \tilde{u}_i = 0$ that
\begin{align}
u_{k,1}(x_2=\pm l/2) &= 0\text{ ,}\label{rb0}\\
\partial_2 u_{k,2}(x_2=\pm l/2) &= 0 \text{ ,}
\end{align}
for the case of non-conducting, and
\begin{align}
\partial_2 u_{k,1}(x_2=\pm l/2) &= 0\text{ ,}\\
\partial_2^2 u_{k,2}(x_2=\pm l/2) &= 0 \text{ ,}\label{fb1}
\end{align}
for the case of transversal-conducting boundaries on top of $u_{k,2}(x_2=\pm l/2) = 0$, for both conditions. For the temperature perturbation one needs to 
require $\theta_k(x_2=\pm l/2) = 0$, since the system is in contact with two heat baths, i.e. its temperature is fixed at the boundaries. 
We can conclude that the transition from a stable to an unstable situation for our set of equations occurs exactly at $s_k = 0$. As a result, 
the critical relativistic Rayleigh number and wave number for non-conducting boundaries at $x_2=\pm l/2$ are approximately given by
\begin{align}
\mathrm{Ra}_c &\approx 1707.762\text{ ,}\label{CrRaRig}\\
k_c &\approx \frac{3.117}{l}\text{ ,}\label{CrWaRig}
\end{align}
which coincides with the classical value \cite{chandra61}. The reason for this is the assumption of a non-relativistic current flow, 
which will lead in linear stability analysis to similar equations as in the classical case, cf. Refs. \cite{chandra61,severin01}, 
if the role of mass and charge are interchanged. Therefore, one should find the same critical Rayleigh number. 
The ultra-relativistic nature of the quasi-particle and the Fermi-Dirac statistics become apparent in the formulae for the constituents 
of the relativistic Rayleigh number, such as thermal charge expansion coefficient, charge density, thermal conductivity, etc.
For transversal-conducting boundaries, the critical Rayleigh number and wave number are given by
\begin{align}
\mathrm{Ra}_c &= \frac{27\pi^4}{4} \approx 657.51 \text{ ,}\label{CrRaFree}\\
k_c &= \frac{\pi}{\sqrt{2}l} \approx \frac{2.221}{l}\text{ .}\label{CrWaFree}
\end{align}
Having derived the relativistic Rayleigh number, and its critical value for the onset of the instability, 
we will now replace the physical values for graphene and see if under actual experimental conditions, it is possible to observe the instability.

\subsection{Application to Graphene}
Despite its high thermal conductivity, $\kappa_R/d \sim 10^{22} \text{ eV/(K\;m\;s)}$, cf. Ref.~\cite{lee11}, 
with $d\sim 10^{-10}\text{ m}$ being the nominal thickness of graphene, its electronic fluid also has a very low shear viscosity~\cite{mueller09},
\begin{equation*}
\eta_R \approx C_{\eta}\hbar \left(\frac{k_B T_R}{\alpha \hbar c}\right)^2 \lbrace 1+\mathcal{O}\left[\log^{-1}(\alpha)\right]\rbrace \sim 10^{-3} \text{ eV s/m}^2 \text{,}
\end{equation*}
where $C_{\eta} \approx 0.449$ is a constant, $\alpha\equiv e^2/\left(\hbar c \varepsilon_r\right) \simeq 2.2/\varepsilon_r$ is the fine structure constant 
and $\varepsilon_r$ is the relative permittivity. Using Eq. (12) in Ref.~\cite{sheehy07} we find for the heat capacity at constant volume 
$\left(c_V\right)_R \sim 10^{10}\text{ eV}/\left(\text{K m}^2\right)$. In the hydrodynamic regime of transport, we deal with a ``Dirac liquid'', 
i.e. we require $n_R \sim 10^{13} \text{ m}^{-2}$, cf. Ref.~\cite{sheehy07}. Therefore, to induce that charge density in the graphene sample, 
we apply an electric field given by $eE_R l \sim 10^{-1} \text{ eV}$, such that the Fermi level changes accordingly. 
In addition, we approximate $\left(\alpha_q\right)_R\beta_R l \approx \frac{T_+-T_-}{T_R} \sim 10^{-2}$. 
Thus, we can make a crude estimate for the relativistic Rayleigh number of graphene with a length $l \sim 100 \mu\text{m}$ 
and $\varepsilon_r \approx 1$ at $T_R \approx 100 \text{ K}$ to find
\begin{equation}
\mathrm{Ra}_{\text{rel}}^{(gr)} \sim 10^3 \text{ .}
\end{equation}
This magnitude is comparable to the critical Rayleigh number calculated in the previous paragraph. 
It tells us that the experimental realization is challenging due to the length of about $100 \mu\text{m}$ 
but probably achievable in the near future, since single-layer samples with sizes larger than $70\mu\text{m}$ have already been produced \cite{sun10,balan10}.

We want to point out that for higher temperatures smaller sample sizes could be used, since the ratio $\left(c_V\right)_R/\kappa_R \sim \mathcal{O}(T^2)$ 
and $eE_R l \sim \mathcal{O}(T)$ while the remaining parameter combination in the relativistic Rayleigh number is almost unaffected. 
However electron-phonon interactions will become more pronounced and hence the dynamics could be altered, cf. Refs. \cite{mendoza13c,bao09}.
\section{Relativistic lattice Boltzmann simulation}
\label{Sim}
\begin{figure}[htb]
\begin{center}
\includegraphics[width=0.45\textwidth]{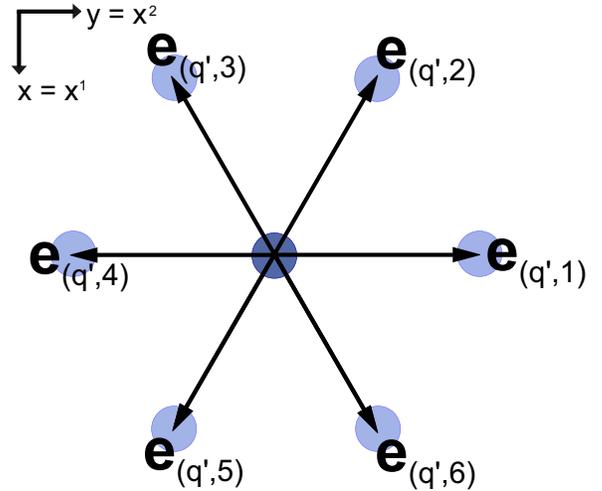}
\caption{Hexagonal lattice structure according to Ref. \cite{oettinger13}}
\label{hexlat}
\end{center}
\end{figure}

In order to test our theoretical predictions we use the method described in Ref. \cite{oettinger13} to perform a relativistic lattice 
Boltzmann simulation of the \textit{electrons} in graphene for vanishing reduced chemical potential. 
The collision operator for the particles in Eq.~\eqref{RBE} is approximated by an Anderson-Witting collision operator \cite{anderson74}, 
i.e. the relativistic Boltzmann equation is modified to
\begin{equation}
p^{\alpha}\frac{\partial f}{\partial x^{\alpha}} + K^{\alpha}\frac{\partial f}{\partial p^{\alpha}} = -\frac{p_{\alpha}U_L^{\alpha}}{c^2 \tau}\left[f-f_{eq}\right]\text{ ,}\label{RBS}
\end{equation}
with $\tau$ the relaxation time and $U_L^{\alpha}$ being the 3-velocity of the fluid element in Landau-Lifshitz decomposition, 
i.e. the 3-velocity of the energy-flow, as opposed to our theoretical treatment where we measured with respect to the charge-flow. 
For the equilibrium distribution we use the ultra-relativistic Fermi-Dirac distribution 
\begin{equation*}
f_{eq} = \frac{4/h^2}{e^{\frac{U_L^{\alpha}p_{\alpha}}{k_B T}}+1}\text{ ,}
\end{equation*}
which leads to
\begin{align}
n_{eq} &= U_L^{\alpha} \langle p_{\alpha} \rangle_{eq} = \frac{\pi}{12}\left(\frac{k_B T}{\hbar c}\right)^2 \label{EqDen}\text{ ,}\\
\epsilon_{eq} &= U_L^{\alpha} \langle p_{\alpha} p_{\beta} \rangle_{eq} U_L^{\beta} = 2 p_{eq} \text{ ,}\\
p_{eq} &= \frac{3\zeta(3)k_B T}{4\pi}\left(\frac{k_B T}{\hbar c}\right)^2\text{ .}
\end{align}
Then, we perform an expansion into orthogonal polynomials, which are written in appendix A of Ref. \cite{oettinger13}, 
\begin{equation*}
f_{eq}(t,\vec{x},p,\vec{v})=\frac{\pi^2}{e^{\frac{p}{T_0}}+1}\sum_{n,k=0}^\infty a_{\underline{i}}^{(n k)}(t,\vec{x})P_{\underline{i}}^{(n)}\left(\vec{v}\right)F^{(k)}\left(p\right),
\end{equation*}
with $\hbar=k_B=c=e=1$ and $\vec{p}=p\vec{v}=p(\cos\phi,\sin\phi)$ in polar coordinates. $T_0$ is a constant, dimensionless lattice temperature.
The relevant coefficients up to $a_{\underline{i}}^{(22)}$ are calculated in Ref. \cite{oettinger13}. 
To simplify the computation even further the angle $\phi$ and radius $p$ are discretized such that a numerical quadrature of the expansion exactly 
reproduces the zeroth, first and second moment of the equilibrium distribution. The corresponding weights and discrete values for the radii can 
be found in appendix C of Ref. \cite{oettinger13}. For the angular quadrature, one finds the weights $\alpha^{(\phi)}_i = 1/6$ and 
angles $\phi_i =  \pi/2 + (i-1)\pi/3$ with $i \in \{1,2,3,4,5,6\}$, leaving us with a hexagonal unit cell for each radius, see Fig. \ref{hexlat}. 
The obtained lattice site distribution function now evolves in time steps $\delta t=\delta x/c$ according to the lattice Boltzmann algorithm with
\begin{align*}
f_{\vec{q}}(t+\delta t,\vec{x}+\hat{e}_{\vec{q}} \delta t)-f_{\vec{q}}(t,\vec{x}) &= \\
-&\frac{p^{\alpha}U_{\alpha}}{p^0 \tau}\left[f_{\vec{q}}(t,\vec{x})-f_{\vec{q}}^{(eq)}(t,\vec{x})\right]
\end{align*}
where $\vec{q} \equiv (q',q'')$ labels the discrete momenta (radius, angle) and $\hat{e}_{\vec{q}} \equiv \vec{v}_{q''} = \vec{p}_{\vec{q}}/p_0$.

\subsection{External Force Implementation}
\begin{figure}[htb]
\begin{center}
\includegraphics{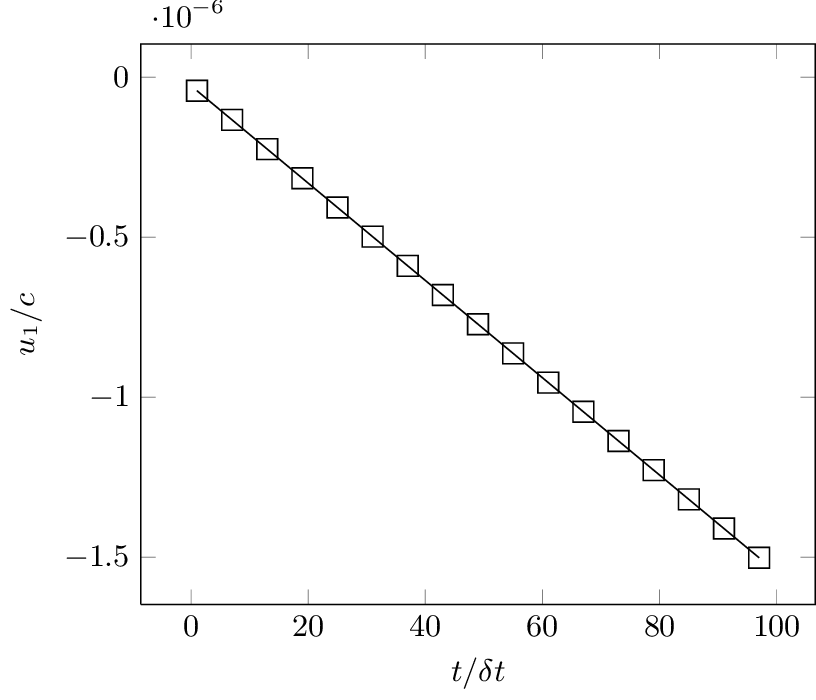}
\caption{First particle-velocity component $u_1$ as a function of time for $E_1=10^{-7}$, $E_2=0$, $T_0=2$, $\tau=\delta t$, 
system size of $64\times 64$, and periodic boundaries. 
The solid line represents the fit: $\frac{u_1(t)}{c} = \left(\frac{10\pi^2}{54 \zeta(3)}\frac{t}{\delta t}-2.69\right) \times 10^{-8}$.}
\label{exF}
\end{center}
\end{figure}

For including the external force in the simulation, we use the same polynomials as in the expansion of the equilibrium distribution to write 
\begin{equation*}
m K^{\alpha}\frac{\partial f}{\partial p^{\alpha}} = \frac{1}{e^{\frac{p}{T_0}}+1}\sum_{n,k=0}^{\infty}b_{\underline{i}}^{(nk)}P_{\underline{i}}^{(n)}(\vec{v})F^{(k)}(p)\text{ .}
\end{equation*}
In addition, Eqs.~\eqref{PC} and \eqref{EMC} need to be fulfilled, i.e.
\begin{align*}
\int \!\frac{\mathrm{d}\vec{p}}{p^0} \hspace{1ex}p^{\beta}m K^{\alpha}\frac{\partial f}{\partial p^{\alpha}}&=-F^{\beta\nu}N_{\nu}\text{ ,}\\
\int \!\frac{\mathrm{d}\vec{p}}{p^0} \hspace{1ex}m K^{\alpha}\frac{\partial f}{\partial p^{\alpha}}&=0 \text{ .}
\end{align*}
From this we infer
\begin{align*} 
-F^{0\nu}N_{\nu}&= T_0^2 \left[\Gamma_F^{(1)} b^{(01)} + c_{10} \Gamma_F^{(0)} b^{(00)}\right]\text{ ,}\\
-F^{i\nu}N_{\nu}&= \frac{T_0^2}{2} \left[\Gamma_F^{(1)} b_i^{(11)} + c_{10} \Gamma_F^{(0)} b_i^{(10)}\right]\text{ ,}\\
0 &= T_0 \Gamma_F^{(0)} b^{(00)}\text{ ,}
\end{align*}
and finally choose
\begin{align}
b^{(01)} &= \frac{<p^i> E_i}{\Gamma_F^{(1)} T_0^2}\text{ ,}\label{coef1}\\
b_i^{(10)} &= \frac{2 <p^0> E_i}{T^2_0 c_{10} \Gamma_F^{(0)}}\text{ ,}\label{coef2}
\end{align}
with all other coefficients equal to zero. For the validation we measure the velocities $u_i \equiv \frac{\langle p^i\rangle}{\langle p^0\rangle}$ 
of the particle 3-flow as function of time for very small fields, such that according to Eq.~\eqref{MC1} a linear relation between time and velocity is expected. 
Choosing a homogeneous temperature $T=T_0$, constant electric field in $x$-direction and zero energy flux as initial conditions, 
the result for the first particle-velocity component as shown in Fig. \ref{exF} confirms the linear dependence. The fit has been done with the function
\begin{equation}
u_1(t) = -\frac{n E_1}{\epsilon+p}t+b\text{ ,}
\end{equation}
where $b$ is a fit parameter. The fit parameter is not zero, since we impose a vanishing energy flux but measure the charge-velocity. 
These velocities differ since the electric field creates a heat flux, e.g. Joule heating. 
We also find that the force does not have any influence on the second velocity component, the density or the pressure, 
i.e. the absolute value of their temporal change always stays below the numerical error of $10^{-16}$. 

\subsection{Shear Viscosity Measurement}

\begin{figure}[htb]
\begin{center}
\includegraphics{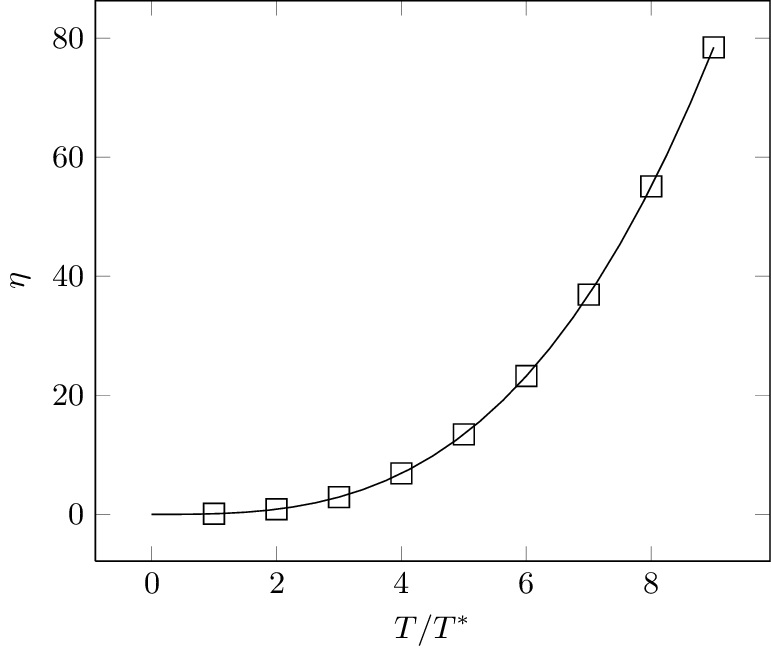}
\caption{Shear viscosity $\eta$ as a function of the temperature with $T^*=100\text{ K}$. 
Here we have used $E_1=10^{-8}$, $E_2=0$, $\tau=\delta t$, 
and a system size of $64\times 64$. The solid line denotes the curve $\eta(x) = \frac{9\zeta(3)x^3}{32\pi}$.}
\label{vt1}
\includegraphics{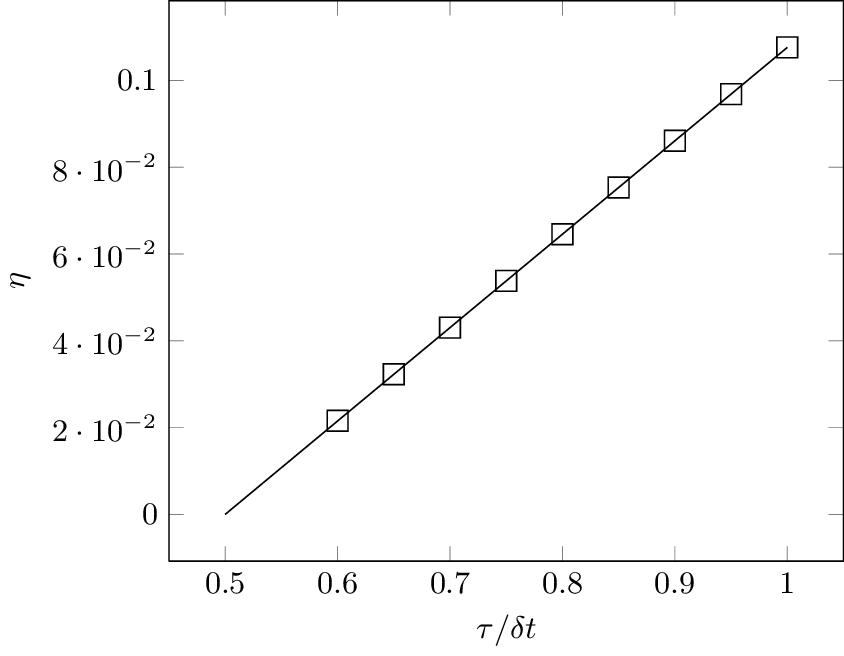}
\caption{Shear viscosity as function of the relaxation time $\tau$. Here we have used $E_1=10^{-8}$, $E_2=0$, $T_0=1$, 
and a system size of $64\times 64$. The solid line represents the curve $\eta(x)=\frac{9\zeta(3)}{16\pi}(x-0.5)$.}
\label{vt2}
\end{center}
\end{figure}

In order to measure the shear viscosity we use a Poiseuille flow test. For this purpose, we examine the \textit{steady} ($t\geq 10^5 \delta t$) velocity profile obtained 
by imposing simple bounce-back boundary conditions in $y$-direction, periodic boundaries in $x$-direction, and applying a small, constant, homogeneous force density 
$\vec{f}=-n E_1 \hat{e}_1$. The initial state is chosen with $T=T_0$ and zero energy flux. As we arrive to the steady state according to Eq.~\eqref{PC1}, 
we know that $\vec{\nabla}\cdot\vec{u}= 0$. Hence we can write for Eq.~\eqref{MC1},
\begin{equation*}
\eta \Delta \vec{u} = n E_1\hat{e}_1\text{ .}
\end{equation*}
As our system is translationally invariant in $x$-direction and we assume that $\eta$ is homogeneous, we expect the velocity to depend only on $y$, i.e.
\begin{equation*}
\eta u_1'' = n E_1\text{ .}
\end{equation*}
Considering the no-slip boundary condition in $y$-direction with height $l$ the velocity profile is
\begin{equation*}
u_1(y) = \frac{n E_1}{2\eta}\left(y^2-l y\right)\text{ ,}
\end{equation*}
which could be confirmed in the measurement. By fitting our results to this profile we are able to extract information about the shear viscosity. 
Thus, we measure the dependence of the shear viscosity on the temperature $T$ and the relaxation time $\tau$, 
which are shown in Figs. \ref{vt1} and \ref{vt2} together with the curve of the function $\eta = \frac{\epsilon+p}{4}\left(\tau-\delta t/2\right)$.
For the conversion to a physical temperature one uses equation \eqref{EqDen} to find the scale $\frac{\hbar c}{k_B \delta x}=T^*$. 
By choosing $\delta x \sim 0.1 \mu\text{m}$ we find $T^* \sim 100\text{ K}$. We have chosen $100\text{ K}$ as a reference temperature to still simulate
the hydrodynamic regime of transport in graphene and at the same time avoid significant effects of the electron-phonon interactions, cf. Refs. \cite{mendoza13c,bao09}.

Note that rigid boundaries stem from a classical understanding of fluid dynamics, 
since in relativity the concept of solid walls is not well defined due to the Lorentz's contraction. 
However, this effect is negligible as we are working in a reference frame with non-relativistic velocities and stationary walls.

\subsection{Thermal Conductivity Measurement}
\label{TCon}
\begin{figure}[htb]
\begin{center}
\includegraphics{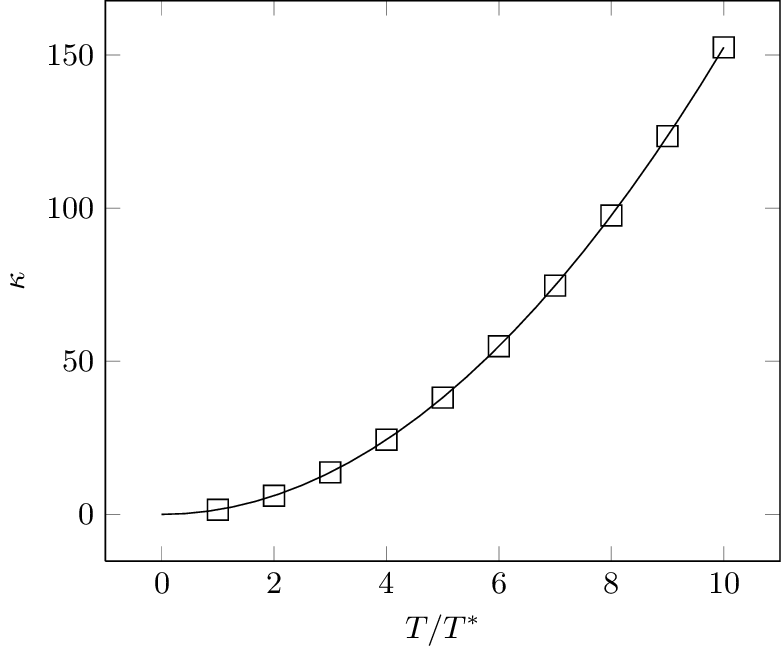}
\caption{Thermal conductivity $\kappa$ as a function of the temperature at $t=10^3 \delta t$ with $T^*=100\text{ K}$. 
The following values have been used: $E_1=10^{-8}$, $E_2=0$, $\tau=\delta t$, and a system size of $32\times 32$. 
The solid line is the fit: $\kappa(x)=1.525 x^2$.}
\label{hc1}
\includegraphics{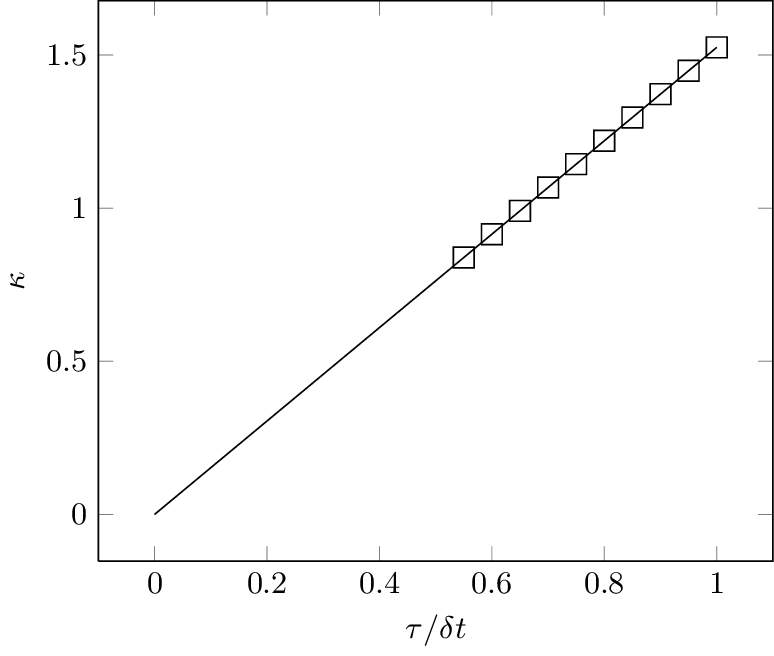}
\caption{Thermal conductivity as a function of the relaxation time $\tau$ at $t=1000\delta t$. 
The following values have been used: $E_1=10^{-6}$, $E_2=0$, $T_0=1$, and a system size of $32\times 32$. 
The solid line denotes the curve given by the expression: $\kappa(x)=1.525 x$.}
\label{hc2}
\end{center}
\end{figure}

According to Ref.~\cite{kremer02} we can measure the heat-flux for processes close to equilibrium by considering the difference in the 3-velocities for 
the Eckart $U^{\alpha}$ and Landau-Lifschitz decomposition $U_L^{\alpha}$, i.e.
\begin{equation}
q^{\alpha} = (\epsilon+p)\left(U_L^{\alpha}-\frac{\langle p^{\alpha}\rangle}{n}\right)\text{ .}\label{lanEck}
\end{equation}
For the measurement of the thermal conductivity, we use periodic boundary conditions in both directions and introduce a very small constant, homogeneous electric field 
into the system. As initial configuration, we choose a homogeneous temperature $T=T_0$ and zero energy flux. Therefore, we are dealing with a homogeneous 
situation, i.e. all spatial gradients vanish and a constant, homogeneous acceleration $\partial_t \vec{u} = -\frac{n\vec{E}}{\epsilon+p}$ is acting on the fluid components. 
According to the constitutive Eq.~\eqref{hflux} the heat flux can be written as
\begin{align*}
\vec{q} &= -\gamma \kappa T \partial_t \begin{pmatrix}
\gamma u_1\\
\gamma u_2 
\end{pmatrix}
\approx \frac{\kappa n T}{\epsilon + p} \begin{pmatrix}
E_1\\
E_2
\end{pmatrix}\text{ ,}
\end{align*}
where for the last approximation we have assumed a non-relativistic charge flow, i.e. $|\vec{u}|/c\ll 1$. 
We find a constant, homogeneous heat flux $\vec{q}$. The temperature and relaxation time dependence of 
the heat conductivity are shown in Figs. \ref{hc1} and \ref{hc2}. Expectedly, we cannot confirm $\kappa \propto \tau -\delta t/2$, 
but instead, we have found $\kappa \approx 1.525\tau T^2$. This is due to the fact that our model recovers 
only the first three moments of the equilibrium distribution, which are not sufficient to match all transport coefficients, 
in particular the thermal conductivity, given by theoretical predictions \cite{mendoza13a}. In the classical two-dimensional lattice Boltzmann 
simulation with a standard Bhatnager-Gross-Krook collision-operator \cite{bhat54} one observes a similar effect when 
using the energy-conserving method from Ref. \cite{karlin05a}, if the moments do not match up to fourth order.

\subsection{Stability Analysis}

\begin{figure}[htb]
\begin{center}
\includegraphics{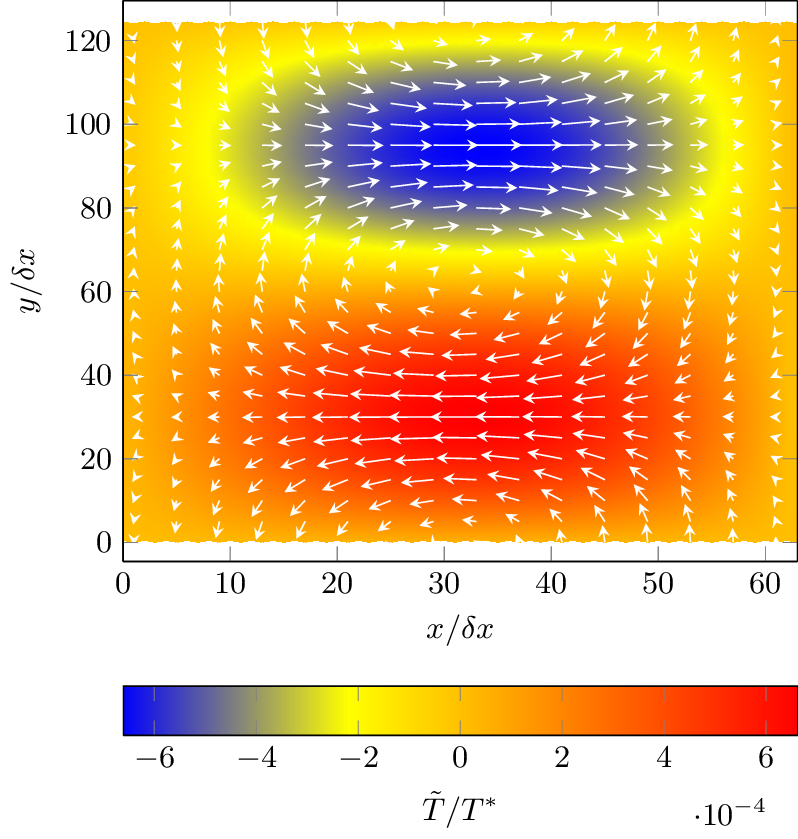}
\caption{Electron velocity $\vec{u}/c$ and temperature perturbation field $\tilde{T}$ at $t=10^5\delta t$ with $T^*=100\text{ K}$. 
The simulation has been performed using a relaxation time $\tau=\delta t$, lattice temperature $T_0=1$, 
$b_0=d_0\approx 4.2\times 10^{-3} \Rightarrow \beta_R > 0$, Rayleigh number $\mathrm{Ra}_{\text{rel}}=1800$, and a system size of $64\times 128$. 
Here $|\vec{u}/c|_{max} \sim 10^{-4}$. }
\label{instab}
\end{center}
\end{figure}

In analogy to the classical lattice Boltzmann simulations of the Rayleigh-B\'enard instability~\cite{shan97}, we include the unperturbed force density 
$\bar{n}e\bar{E}_1 \hat{e}_1=-\vec{\nabla}V$ into the pressure term, i.e. $p \rightarrow p + V$, and could start the simulation with a \textit{homogeneous} temperature 
distribution $T=T_0$ and zero energy flux. Since triggering an instability can be very complicated in numerical simulations, we have based our triggering mechanism on 
the method used in Ref.~\cite{shan97}. Following this work, we proceed by changing the coefficients of the external force in Eqs. \eqref{coef1} and \eqref{coef2} by
\begin{align*}
b^{(01)} &= \frac{b_0 \langle p^1\rangle}{\langle p^0 \rangle\Gamma_F^{(1)} T_0^2}\text{ ,}\\
b_i^{(10)} &= \frac{2d_0(T/T_0-1)}{c_{10} \Gamma_F^{(0)}T^2_0} \delta_{i1}\text{ ,}
\end{align*}
where the temperature distribution is calculated by
\begin{equation*}
T = \frac{\pi^2}{18\zeta(3)}\frac{\epsilon}{n}\text{ .}
\end{equation*}
The new parameters stand for $b_R=\left(c_V\right)_R\beta_R$, and $d_R=\rho_R\left(\alpha_q\right)_R a_R$. Thus, the relativistic Rayleigh number reads
\begin{equation}
\mathrm{Ra}_{\text{rel}} = \frac{d_R b_R l^4}{\kappa_R \eta_R}\approx 3.047\frac{b_0 d_0 (l/\delta x)^4}{T_0^5 \tau/\delta t(\tau/\delta t-1/2)} \text{ .} \label{CrRaSim}
\end{equation}
Since, in our simulations, we are using the Landau-Lifschitz decomposition, it is not straightforward to impose the non- or transversal-conducting boundaries 
from Eqs. \eqref{rb0} to \eqref{fb1}, and therefore, we approximate the non-conducting thermal boundaries by replacing the distribution function with the equilibrium 
value for $T=T_0$ and $\vec{u}_L = 0$ at the edges in $x$-direction. In $y$-direction we choose periodic boundaries. The height of the system is chosen twice as long 
as the width, since according to Eq.~\eqref{CrWaRig} the critical wave length is $\lambda_c \approx 2.016 l$. As initial condition, we start with a temperature distribution 
\begin{equation*}
T(x,y,t=0) = T_0\left[1+10^{-8}\frac{4x}{l}\left(1-\frac{x}{l}\right)\cos(k_c y)\right]\text{ .}
\end{equation*}
One observes the expected formation of convection cells with a wave length $\lambda \approx 2 l$ 
and a cosine-shaped temperature perturbation, which vanishes at the thermal contacts, as shown in Fig.~\ref{instab}.

\begin{figure}[htb]
\begin{center}
\includegraphics{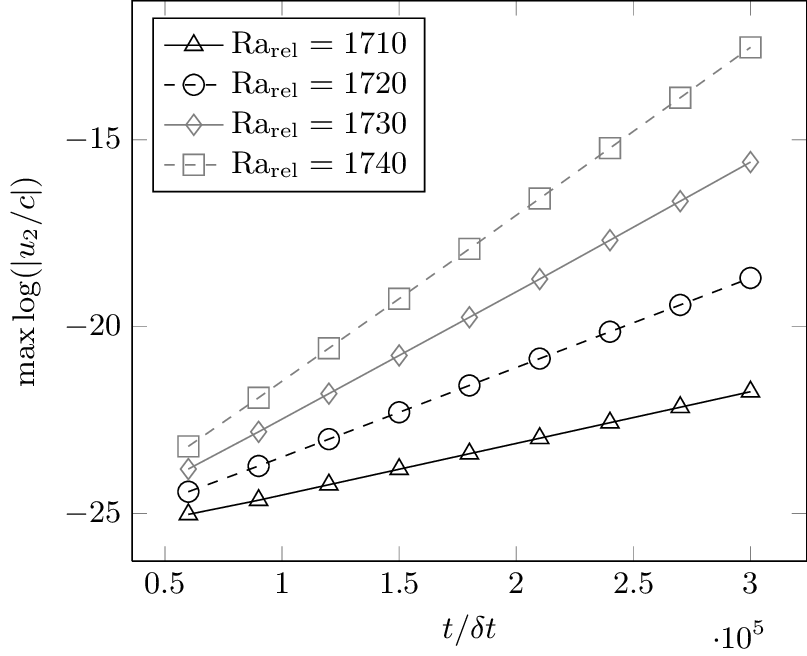}
\caption{Logarithm of the peak vertical velocity as a function of time for a system size of $64\times 128$}
\label{growth}
\includegraphics{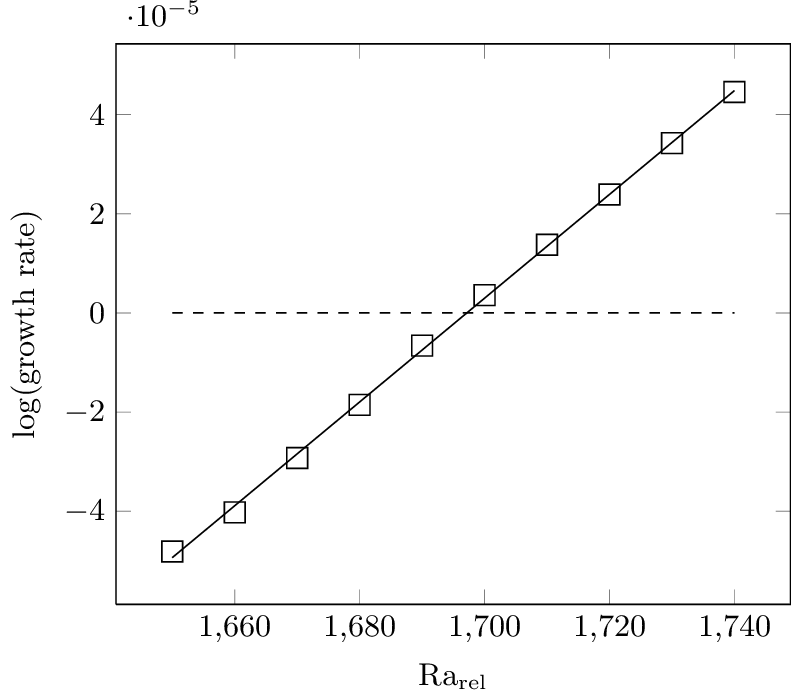}
\caption{Logarithmic growth rate as a function of the relativistic Rayleigh number for a system size of $64\times 128$ with dashed horizontal line at $y=0$}
\label{RayCrit}
\end{center}
\end{figure}

As a next step, we use Eq.~\eqref{CrRaSim} to identify the critical Rayleigh number of our setup and compare it to our theoretical value, given by Eq.~\eqref{CrRaRig}. 
This is done by measuring the logarithmic growth rate of the peak velocity of the charge flow perpendicular to the applied electric field $u_2$ at near-critical Rayleigh numbers, 
similar to the procedure described in Ref. \cite{shan97}. The maximum velocity is measured for $ 10^{-4}t\in \{6 \delta t,9\delta t,\dots,30\delta t\}$. In Fig.~\eqref{growth} we see the exponential growth. 
In Fig.~\ref{RayCrit} we observe the linear dependence and measure a critical Rayleigh number of $\mathrm{Ra}_c \approx 1697$, which is about $0.6\%$ smaller than the theoretical value in Eq.~\eqref{CrRaRig}. 
This shows that our numerical results have excellent agreement with our theoretical predictions. 

\section{Conclusion}
\label{Concl}
We studied both, theoretically and numerically, the relativistic Rayleigh-B\'enard instability for the charge transport in quantum critical graphene. 
As a new driving mechanism we introduced an electrical field to mimic the effects of the buoyancy force in the classical phenomenon, 
i.e. interchanging the role of mass and charge. We have found that the classical inhibitors (viscous forces and heat transport) and 
drivers (temperature gradient and external force) keep their role, and the instability can develop. Using a linear stability analysis, 
we derived a quantitative criterion ($\mathrm{Ra}_{\text{rel}}\geq \text{Ra}_c$) for the occurrence of the instability which compares 
the relativistic Rayleigh number of the system with the critical values in Eqs. ~\eqref{CrRaRig} and \eqref{CrRaFree}, whose magnitude
depends on the nature of the boundary conditions. The ultra-relativistic fermionic plasma influences the formula for the relativistic Rayleigh number, 
but under the assumption of non-relativistic current flow it does not alter the critical Rayleigh number. 
Applying our result to graphene we have found that the experimental realization for 
the appearance of convection cells is challenging due to a large sample length of about $100\mu\text{m}$ but probably achievable in the 
near future. The reason for the large sample length are twofold. On one hand graphene's high thermal conductivity supresses the instability.
On the other hand we are constrained by the hydrodynamic regime of transport which limits the value for the charge density assuming a given average temperature
where electron-electron and -hole collisions dominate over other scattering mechanisms.
For the numerical simulations we have improved the method proposed in Ref.~\cite{oettinger13}, by adding an external force and 
determined the functional dependence of the transport coefficients on temperature and relaxation time. We have demonstrated the 
occurrence of the convection cell pattern with the expected critical wavelength and temperature distributions in the case of non-conducting boundaries. 
In addition, we have measured a critical relativistic Rayleigh number which is in very good agreement to our theoretical predictions from the stability analysis.
Extending our theoretical results by analyzing non-Boussinesq effects, studying the influence of substrates, as e.g. $\mathrm{SiC}$, considering 
electron-phonon interactions on the instability, as well as applying our finding to thermo-electronic devices, will be a subject of future research. 

\begin{acknowledgments}
Financial support from the European Research Council (ERC) Advanced Grant 319968-FlowCCS is kindly acknowledged. I.K. was supported by the European Research Council (ERC) Advanced Grant No. 291094-ELBM.
\end{acknowledgments}

\bibliography{lit}{}

\begin{thebibliography}{37}%
\makeatletter
\providecommand \@ifxundefined [1]{%
 \@ifx{#1\undefined}
}%
\providecommand \@ifnum [1]{%
 \ifnum #1\expandafter \@firstoftwo
 \else \expandafter \@secondoftwo
 \fi
}%
\providecommand \@ifx [1]{%
 \ifx #1\expandafter \@firstoftwo
 \else \expandafter \@secondoftwo
 \fi
}%
\providecommand \natexlab [1]{#1}%
\providecommand \enquote  [1]{``#1''}%
\providecommand \bibnamefont  [1]{#1}%
\providecommand \bibfnamefont [1]{#1}%
\providecommand \citenamefont [1]{#1}%
\providecommand \href@noop [0]{\@secondoftwo}%
\providecommand \href [0]{\begingroup \@sanitize@url \@href}%
\providecommand \@href[1]{\@@startlink{#1}\@@href}%
\providecommand \@@href[1]{\endgroup#1\@@endlink}%
\providecommand \@sanitize@url [0]{\catcode `\\12\catcode `\$12\catcode
  `\&12\catcode `\#12\catcode `\^12\catcode `\_12\catcode `\%12\relax}%
\providecommand \@@startlink[1]{}%
\providecommand \@@endlink[0]{}%
\providecommand \url  [0]{\begingroup\@sanitize@url \@url }%
\providecommand \@url [1]{\endgroup\@href {#1}{\urlprefix }}%
\providecommand \urlprefix  [0]{URL }%
\providecommand \Eprint [0]{\href }%
\providecommand \doibase [0]{http://dx.doi.org/}%
\providecommand \selectlanguage [0]{\@gobble}%
\providecommand \bibinfo  [0]{\@secondoftwo}%
\providecommand \bibfield  [0]{\@secondoftwo}%
\providecommand \translation [1]{[#1]}%
\providecommand \BibitemOpen [0]{}%
\providecommand \bibitemStop [0]{}%
\providecommand \bibitemNoStop [0]{.\EOS\space}%
\providecommand \EOS [0]{\spacefactor3000\relax}%
\providecommand \BibitemShut  [1]{\csname bibitem#1\endcsname}%
\let\auto@bib@innerbib\@empty
\bibitem [{\citenamefont {Novoselov}\ \emph {et~al.}(2004)\citenamefont
  {Novoselov}, \citenamefont {Geim}, \citenamefont {Morozov}, \citenamefont
  {Jiang}, \citenamefont {Zhang}, \citenamefont {Dubonos}, \citenamefont
  {Grigorieva},\ and\ \citenamefont {Firsov}}]{geim04}%
  \BibitemOpen
  \bibfield  {author} {\bibinfo {author} {\bibfnamefont {K.~S.}\ \bibnamefont
  {Novoselov}}, \bibinfo {author} {\bibfnamefont {A.~K.}\ \bibnamefont {Geim}},
  \bibinfo {author} {\bibfnamefont {S.~V.}\ \bibnamefont {Morozov}}, \bibinfo
  {author} {\bibfnamefont {D.}~\bibnamefont {Jiang}}, \bibinfo {author}
  {\bibfnamefont {Y.}~\bibnamefont {Zhang}}, \bibinfo {author} {\bibfnamefont
  {S.~V.}\ \bibnamefont {Dubonos}}, \bibinfo {author} {\bibfnamefont {I.~V.}\
  \bibnamefont {Grigorieva}}, \ and\ \bibinfo {author} {\bibfnamefont {A.~A.}\
  \bibnamefont {Firsov}},\ }\href@noop {} {\bibfield  {journal} {\bibinfo
  {journal} {Science}\ }\textbf {\bibinfo {volume} {306}} (\bibinfo {year}
  {2004})}\BibitemShut {NoStop}%
\bibitem [{\citenamefont {Novoselov}\ \emph {et~al.}(2005)\citenamefont
  {Novoselov}, \citenamefont {Geim}, \citenamefont {Morozov}, \citenamefont
  {Jiang},\ and\ \citenamefont {Firsov}}]{geim05}%
  \BibitemOpen
  \bibfield  {author} {\bibinfo {author} {\bibfnamefont {K.}~\bibnamefont
  {Novoselov}}, \bibinfo {author} {\bibfnamefont {A.}~\bibnamefont {Geim}},
  \bibinfo {author} {\bibfnamefont {S.}~\bibnamefont {Morozov}}, \bibinfo
  {author} {\bibfnamefont {D.}~\bibnamefont {Jiang}}, \ and\ \bibinfo {author}
  {\bibfnamefont {A.}~\bibnamefont {Firsov}},\ }\href@noop {} {\bibfield
  {journal} {\bibinfo  {journal} {Nature (London)}\ }\textbf {\bibinfo {volume}
  {438}} (\bibinfo {year} {2005})}\BibitemShut {NoStop}%
\bibitem [{\citenamefont {Müller}\ and\ \citenamefont
  {Sachdev}(2008)}]{mueller08a}%
  \BibitemOpen
  \bibfield  {author} {\bibinfo {author} {\bibfnamefont {M.}~\bibnamefont
  {Müller}}\ and\ \bibinfo {author} {\bibfnamefont {S.}~\bibnamefont
  {Sachdev}},\ }\href@noop {} {\bibfield  {journal} {\bibinfo  {journal}
  {Physical Review B}\ }\textbf {\bibinfo {volume} {78}} (\bibinfo {year}
  {2008})}\BibitemShut {NoStop}%
\bibitem [{\citenamefont {Müller}\ \emph {et~al.}(2008)\citenamefont {Müller},
  \citenamefont {Fritz},\ and\ \citenamefont {Sachdev}}]{mueller08b}%
  \BibitemOpen
  \bibfield  {author} {\bibinfo {author} {\bibfnamefont {M.}~\bibnamefont
  {Müller}}, \bibinfo {author} {\bibfnamefont {L.}~\bibnamefont {Fritz}}, \
  and\ \bibinfo {author} {\bibfnamefont {S.}~\bibnamefont {Sachdev}},\
  }\href@noop {} {\bibfield  {journal} {\bibinfo  {journal} {Physical Review
  B}\ }\textbf {\bibinfo {volume} {78}} (\bibinfo {year} {2008})}\BibitemShut
  {NoStop}%
\bibitem [{\citenamefont {Fritz}\ \emph {et~al.}(2008)\citenamefont {Fritz},
  \citenamefont {Schmalian}, \citenamefont {Müller},\ and\ \citenamefont
  {Sachdev}}]{fritz08}%
  \BibitemOpen
  \bibfield  {author} {\bibinfo {author} {\bibfnamefont {L.}~\bibnamefont
  {Fritz}}, \bibinfo {author} {\bibfnamefont {J.}~\bibnamefont {Schmalian}},
  \bibinfo {author} {\bibfnamefont {M.}~\bibnamefont {Müller}}, \ and\ \bibinfo
  {author} {\bibfnamefont {S.}~\bibnamefont {Sachdev}},\ }\href@noop {}
  {\bibfield  {journal} {\bibinfo  {journal} {Phys. Rev. B}\ }\textbf {\bibinfo
  {volume} {78}} (\bibinfo {year} {2008})}\BibitemShut {NoStop}%
\bibitem [{\citenamefont {Müller}\ \emph {et~al.}(2009)\citenamefont {Müller},
  \citenamefont {Schmalian},\ and\ \citenamefont {Fritz}}]{mueller09}%
  \BibitemOpen
  \bibfield  {author} {\bibinfo {author} {\bibfnamefont {M.}~\bibnamefont
  {Müller}}, \bibinfo {author} {\bibfnamefont {J.}~\bibnamefont {Schmalian}}, \
  and\ \bibinfo {author} {\bibfnamefont {L.}~\bibnamefont {Fritz}},\
  }\href@noop {} {\bibfield  {journal} {\bibinfo  {journal} {Physical Review
  Letters}\ }\textbf {\bibinfo {volume} {103}} (\bibinfo {year}
  {2009})}\BibitemShut {NoStop}%
\bibitem [{\citenamefont {Kovtun}\ \emph {et~al.}(2005)\citenamefont {Kovtun},
  \citenamefont {Son},\ and\ \citenamefont {Starinets}}]{kovtun05}%
  \BibitemOpen
  \bibfield  {author} {\bibinfo {author} {\bibfnamefont {P.~K.}\ \bibnamefont
  {Kovtun}}, \bibinfo {author} {\bibfnamefont {D.~T.}\ \bibnamefont {Son}}, \
  and\ \bibinfo {author} {\bibfnamefont {A.~O.}\ \bibnamefont {Starinets}},\
  }\href {\doibase 10.1103/PhysRevLett.94.111601} {\bibfield  {journal}
  {\bibinfo  {journal} {Phys. Rev. Lett.}\ }\textbf {\bibinfo {volume} {94}},\
  \bibinfo {pages} {111601} (\bibinfo {year} {2005})}\BibitemShut {NoStop}%
\bibitem [{\citenamefont {Mendoza}\ \emph {et~al.}(2011)\citenamefont
  {Mendoza}, \citenamefont {Herrmann},\ and\ \citenamefont
  {Succi}}]{mendoza11}%
  \BibitemOpen
  \bibfield  {author} {\bibinfo {author} {\bibfnamefont {M.}~\bibnamefont
  {Mendoza}}, \bibinfo {author} {\bibfnamefont {H.}~\bibnamefont {Herrmann}}, \
  and\ \bibinfo {author} {\bibfnamefont {S.}~\bibnamefont {Succi}},\
  }\href@noop {} {\bibfield  {journal} {\bibinfo  {journal} {Physical Review
  Letters}\ }\textbf {\bibinfo {volume} {106}} (\bibinfo {year}
  {2011})}\BibitemShut {NoStop}%
\bibitem [{\citenamefont {Bénard}(1901)}]{benard01}%
  \BibitemOpen
  \bibfield  {author} {\bibinfo {author} {\bibfnamefont {H.}~\bibnamefont
  {Bénard}},\ }\emph {\bibinfo {title} {{Les Tourbillons cellulaires dans une
  nappe liquide propageant de la chaleur par convection en r\' egime
  permanent}}},\ \href@noop {} {Ph.D. thesis},\ \bibinfo  {school} {Coll\` ege
  de France} (\bibinfo {year} {1901})\BibitemShut {NoStop}%
\bibitem [{\citenamefont {Rayleigh}(1916)}]{rayleigh16}%
  \BibitemOpen
  \bibfield  {author} {\bibinfo {author} {\bibfnamefont {L.}~\bibnamefont
  {Rayleigh}},\ }\href@noop {} {\bibfield  {journal} {\bibinfo  {journal}
  {Philisophical Magazine}\ }\textbf {\bibinfo {volume} {6}},\ \bibinfo {pages}
  {529–546} (\bibinfo {year} {1916})}\BibitemShut {NoStop}%
\bibitem [{\citenamefont {Chandrasekhar}(1961)}]{chandra61}%
  \BibitemOpen
  \bibfield  {author} {\bibinfo {author} {\bibfnamefont {S.}~\bibnamefont
  {Chandrasekhar}},\ }\href@noop {} {\emph {\bibinfo {title} {{Hydrodynamic and
  Hydromagnetic Stability}}}}\ (\bibinfo  {publisher} {Oxford University
  Press},\ \bibinfo {year} {1961})\BibitemShut {NoStop}%
\bibitem [{\citenamefont {Massaioli}\ \emph {et~al.}(1993)\citenamefont
  {Massaioli}, \citenamefont {Benzi},\ and\ \citenamefont {Succi}}]{succi93}%
  \BibitemOpen
  \bibfield  {author} {\bibinfo {author} {\bibfnamefont {F.}~\bibnamefont
  {Massaioli}}, \bibinfo {author} {\bibfnamefont {R.}~\bibnamefont {Benzi}}, \
  and\ \bibinfo {author} {\bibfnamefont {S.}~\bibnamefont {Succi}},\ }\href
  {http://stacks.iop.org/0295-5075/21/i=3/a=009} {\bibfield  {journal}
  {\bibinfo  {journal} {EPL (Europhysics Letters)}\ }\textbf {\bibinfo {volume}
  {21}},\ \bibinfo {pages} {305} (\bibinfo {year} {1993})}\BibitemShut
  {NoStop}%
\bibitem [{\citenamefont {Ahlers}\ \emph {et~al.}(2012)\citenamefont {Ahlers},
  \citenamefont {Bodenschatz}, \citenamefont {Funfschilling}, \citenamefont
  {Grossmann}, \citenamefont {He}, \citenamefont {Lohse}, \citenamefont
  {Stevens},\ and\ \citenamefont {Verzicco}}]{ahlers12}%
  \BibitemOpen
  \bibfield  {author} {\bibinfo {author} {\bibfnamefont {G.}~\bibnamefont
  {Ahlers}}, \bibinfo {author} {\bibfnamefont {E.}~\bibnamefont {Bodenschatz}},
  \bibinfo {author} {\bibfnamefont {D.}~\bibnamefont {Funfschilling}}, \bibinfo
  {author} {\bibfnamefont {S.}~\bibnamefont {Grossmann}}, \bibinfo {author}
  {\bibfnamefont {X.}~\bibnamefont {He}}, \bibinfo {author} {\bibfnamefont
  {D.}~\bibnamefont {Lohse}}, \bibinfo {author} {\bibfnamefont {R.~J. A.~M.}\
  \bibnamefont {Stevens}}, \ and\ \bibinfo {author} {\bibfnamefont
  {R.}~\bibnamefont {Verzicco}},\ }\href {\doibase
  10.1103/PhysRevLett.109.114501} {\bibfield  {journal} {\bibinfo  {journal}
  {Phys. Rev. Lett.}\ }\textbf {\bibinfo {volume} {109}},\ \bibinfo {pages}
  {114501} (\bibinfo {year} {2012})}\BibitemShut {NoStop}%
\bibitem [{\citenamefont {Moon}\ and\ \citenamefont {Chung}(2014)}]{moon14}%
  \BibitemOpen
  \bibfield  {author} {\bibinfo {author} {\bibfnamefont {J.-Y.}\ \bibnamefont
  {Moon}}\ and\ \bibinfo {author} {\bibfnamefont {B.-J.}\ \bibnamefont
  {Chung}},\ }\href {\doibase 10.1016/j.nucengdes.2014.04.017} {\bibfield
  {journal} {\bibinfo  {journal} {Nuclear Engineering and Design}\ }\textbf
  {\bibinfo {volume} {274}},\ \bibinfo {pages} {146–153} (\bibinfo {year}
  {2014})}\BibitemShut {NoStop}%
\bibitem [{\citenamefont {Balandin}\ \emph {et~al.}(2008)\citenamefont
  {Balandin}, \citenamefont {Ghosh}, \citenamefont {Bao}, \citenamefont
  {Calizo}, \citenamefont {Teweldebrhan}, \citenamefont {Miao},\ and\
  \citenamefont {Lau}}]{balandin08}%
  \BibitemOpen
  \bibfield  {author} {\bibinfo {author} {\bibfnamefont {A.~A.}\ \bibnamefont
  {Balandin}}, \bibinfo {author} {\bibfnamefont {S.}~\bibnamefont {Ghosh}},
  \bibinfo {author} {\bibfnamefont {W.}~\bibnamefont {Bao}}, \bibinfo {author}
  {\bibfnamefont {I.}~\bibnamefont {Calizo}}, \bibinfo {author} {\bibfnamefont
  {D.}~\bibnamefont {Teweldebrhan}}, \bibinfo {author} {\bibfnamefont
  {F.}~\bibnamefont {Miao}}, \ and\ \bibinfo {author} {\bibfnamefont {C.~N.}\
  \bibnamefont {Lau}},\ }\href {\doibase 10.1021/nl0731872} {\bibfield
  {journal} {\bibinfo  {journal} {Nano Letters}\ }\textbf {\bibinfo {volume}
  {8}},\ \bibinfo {pages} {902–907} (\bibinfo {year} {2008})}\BibitemShut
  {NoStop}%
\bibitem [{\citenamefont {Öttinger}\ \emph {et~al.}(2013)\citenamefont
  {Öttinger}, \citenamefont {Mendoza},\ and\ \citenamefont
  {Herrmann}}]{oettinger13}%
  \BibitemOpen
  \bibfield  {author} {\bibinfo {author} {\bibfnamefont {D.}~\bibnamefont
  {Öttinger}}, \bibinfo {author} {\bibfnamefont {M.}~\bibnamefont {Mendoza}}, \
  and\ \bibinfo {author} {\bibfnamefont {H.}~\bibnamefont {Herrmann}},\
  }\href@noop {} {\bibfield  {journal} {\bibinfo  {journal} {Physical Review
  E}\ }\textbf {\bibinfo {volume} {88}} (\bibinfo {year} {2013})}\BibitemShut
  {NoStop}%
\bibitem [{\citenamefont {Benzi}\ \emph {et~al.}(1992)\citenamefont {Benzi},
  \citenamefont {Succi},\ and\ \citenamefont {Vergassola}}]{succi92}%
  \BibitemOpen
  \bibfield  {author} {\bibinfo {author} {\bibfnamefont {R.}~\bibnamefont
  {Benzi}}, \bibinfo {author} {\bibfnamefont {S.}~\bibnamefont {Succi}}, \ and\
  \bibinfo {author} {\bibfnamefont {M.}~\bibnamefont {Vergassola}},\ }\href
  {\doibase 10.1016/0370-1573(92)90090-M} {\bibfield  {journal} {\bibinfo
  {journal} {Physics Reports}\ }\textbf {\bibinfo {volume} {222}},\ \bibinfo
  {pages} {145–197} (\bibinfo {year} {1992})}\BibitemShut {NoStop}%
\bibitem [{\citenamefont {Mendoza}\ \emph {et~al.}(2010)\citenamefont
  {Mendoza}, \citenamefont {Boghosian}, \citenamefont {Herrmann},\ and\
  \citenamefont {Succi}}]{mendoza10}%
  \BibitemOpen
  \bibfield  {author} {\bibinfo {author} {\bibfnamefont {M.}~\bibnamefont
  {Mendoza}}, \bibinfo {author} {\bibfnamefont {B.~M.}\ \bibnamefont
  {Boghosian}}, \bibinfo {author} {\bibfnamefont {H.~J.}\ \bibnamefont
  {Herrmann}}, \ and\ \bibinfo {author} {\bibfnamefont {S.}~\bibnamefont
  {Succi}},\ }\href {\doibase 10.1103/PhysRevLett.105.014502} {\bibfield
  {journal} {\bibinfo  {journal} {Phys. Rev. Lett.}\ }\textbf {\bibinfo
  {volume} {105}},\ \bibinfo {pages} {014502} (\bibinfo {year}
  {2010})}\BibitemShut {NoStop}%
\bibitem [{\citenamefont {Uehling}\ and\ \citenamefont
  {Uhlenbeck}(1933)}]{uehling33}%
  \BibitemOpen
  \bibfield  {author} {\bibinfo {author} {\bibfnamefont {E.~A.}\ \bibnamefont
  {Uehling}}\ and\ \bibinfo {author} {\bibfnamefont {G.~E.}\ \bibnamefont
  {Uhlenbeck}},\ }\href {\doibase 10.1103/PhysRev.43.552} {\bibfield  {journal}
  {\bibinfo  {journal} {Phys. Rev.}\ }\textbf {\bibinfo {volume} {43}},\
  \bibinfo {pages} {552–561} (\bibinfo {year} {1933})}\BibitemShut {NoStop}%
\bibitem [{\citenamefont {Cercignani}\ and\ \citenamefont
  {Kremer}(2002)}]{kremer02}%
  \BibitemOpen
  \bibfield  {author} {\bibinfo {author} {\bibfnamefont {C.}~\bibnamefont
  {Cercignani}}\ and\ \bibinfo {author} {\bibfnamefont {G.}~\bibnamefont
  {Kremer}},\ }\href@noop {} {\emph {\bibinfo {title} {{The Relativistic
  Boltzmann Equation: Theory and Application}}}}\ (\bibinfo  {publisher}
  {Birkhauser-Verlag},\ \bibinfo {year} {2002})\BibitemShut {NoStop}%
\bibitem [{\citenamefont {Landau}\ and\ \citenamefont
  {Lifshitz}(1987)}]{landau87}%
  \BibitemOpen
  \bibfield  {author} {\bibinfo {author} {\bibfnamefont {L.}~\bibnamefont
  {Landau}}\ and\ \bibinfo {author} {\bibfnamefont {E.}~\bibnamefont
  {Lifshitz}},\ }\href@noop {} {\emph {\bibinfo {title} {{Course of Theoretical
  Physics}}}},\ \bibinfo {edition} {2nd}\ ed.,\ Vol.~\bibinfo {volume} {6}\
  (\bibinfo  {publisher} {Pergamon Press},\ \bibinfo {year} {1987})\BibitemShut
  {NoStop}%
\bibitem [{\citenamefont {Wangsness}(1986)}]{wang86}%
  \BibitemOpen
  \bibfield  {author} {\bibinfo {author} {\bibfnamefont {R.~K.}\ \bibnamefont
  {Wangsness}},\ }\href@noop {} {\emph {\bibinfo {title} {{Electromagnetic
  Fields}}}},\ \bibinfo {edition} {2nd}\ ed.\ (\bibinfo  {publisher} {Wiley},\
  \bibinfo {year} {1986})\BibitemShut {NoStop}%
\bibitem [{\citenamefont {Severin}\ and\ \citenamefont
  {Herwig}(2001)}]{severin01}%
  \BibitemOpen
  \bibfield  {author} {\bibinfo {author} {\bibfnamefont {J.}~\bibnamefont
  {Severin}}\ and\ \bibinfo {author} {\bibfnamefont {H.}~\bibnamefont
  {Herwig}},\ }\href@noop {} {\bibfield  {journal} {\bibinfo  {journal}
  {Forschung im Ingenieurwesen}\ }\textbf {\bibinfo {volume} {66}} (\bibinfo
  {year} {2001})}\BibitemShut {NoStop}%
\bibitem [{\citenamefont {Seol}\ \emph {et~al.}(2010)\citenamefont {Seol},
  \citenamefont {Jo}, \citenamefont {Moore}, \citenamefont {Lindsay},
  \citenamefont {Aitken}, \citenamefont {Pettes}, \citenamefont {Li},
  \citenamefont {Yao}, \citenamefont {Huang}, \citenamefont {Broido},
  \citenamefont {Mingo}, \citenamefont {Ruoff},\ and\ \citenamefont
  {Shi}}]{seol10}%
  \BibitemOpen
  \bibfield  {author} {\bibinfo {author} {\bibfnamefont {J.~H.}\ \bibnamefont
  {Seol}}, \bibinfo {author} {\bibfnamefont {I.}~\bibnamefont {Jo}}, \bibinfo
  {author} {\bibfnamefont {A.~L.}\ \bibnamefont {Moore}}, \bibinfo {author}
  {\bibfnamefont {L.}~\bibnamefont {Lindsay}}, \bibinfo {author} {\bibfnamefont
  {Z.~H.}\ \bibnamefont {Aitken}}, \bibinfo {author} {\bibfnamefont {M.~T.}\
  \bibnamefont {Pettes}}, \bibinfo {author} {\bibfnamefont {X.}~\bibnamefont
  {Li}}, \bibinfo {author} {\bibfnamefont {Z.}~\bibnamefont {Yao}}, \bibinfo
  {author} {\bibfnamefont {R.}~\bibnamefont {Huang}}, \bibinfo {author}
  {\bibfnamefont {D.}~\bibnamefont {Broido}}, \bibinfo {author} {\bibfnamefont
  {N.}~\bibnamefont {Mingo}}, \bibinfo {author} {\bibfnamefont {R.~S.}\
  \bibnamefont {Ruoff}}, \ and\ \bibinfo {author} {\bibfnamefont
  {L.}~\bibnamefont {Shi}},\ }\href {\doibase 10.1126/science.1184014}
  {\bibfield  {journal} {\bibinfo  {journal} {Science}\ }\textbf {\bibinfo
  {volume} {328}},\ \bibinfo {pages} {213–216} (\bibinfo {year}
  {2010})}\BibitemShut {NoStop}%
\bibitem [{\citenamefont {Herwig}\ and\ \citenamefont
  {Schäfer}(1992)}]{herwig92}%
  \BibitemOpen
  \bibfield  {author} {\bibinfo {author} {\bibfnamefont {H.}~\bibnamefont
  {Herwig}}\ and\ \bibinfo {author} {\bibfnamefont {P.}~\bibnamefont
  {Schäfer}},\ }\href@noop {} {\bibfield  {journal} {\bibinfo  {journal}
  {Journal of Fluid Mechanics}\ }\textbf {\bibinfo {volume} {243}},\ \bibinfo
  {pages} {pp. 1–14} (\bibinfo {year} {1992})}\BibitemShut {NoStop}%
\bibitem [{\citenamefont {Jackson}(1999)}]{jackson99}%
  \BibitemOpen
  \bibfield  {author} {\bibinfo {author} {\bibfnamefont {J.~D.}\ \bibnamefont
  {Jackson}},\ }\href@noop {} {\emph {\bibinfo {title} {{Classical
  Electrodynamics}}}},\ \bibinfo {edition} {3rd}\ ed.\ (\bibinfo  {publisher}
  {Wiley},\ \bibinfo {year} {1999})\BibitemShut {NoStop}%
\bibitem [{\citenamefont {Lee}\ \emph {et~al.}(2011)\citenamefont {Lee},
  \citenamefont {Yoon}, \citenamefont {Kim}, \citenamefont {Lee},\ and\
  \citenamefont {Cheong}}]{lee11}%
  \BibitemOpen
  \bibfield  {author} {\bibinfo {author} {\bibfnamefont {J.-U.}\ \bibnamefont
  {Lee}}, \bibinfo {author} {\bibfnamefont {D.}~\bibnamefont {Yoon}}, \bibinfo
  {author} {\bibfnamefont {H.}~\bibnamefont {Kim}}, \bibinfo {author}
  {\bibfnamefont {S.~W.}\ \bibnamefont {Lee}}, \ and\ \bibinfo {author}
  {\bibfnamefont {H.}~\bibnamefont {Cheong}},\ }\href@noop {} {\bibfield
  {journal} {\bibinfo  {journal} {Phys. Rev. B}\ }\textbf {\bibinfo {volume}
  {83}} (\bibinfo {year} {2011})}\BibitemShut {NoStop}%
\bibitem [{\citenamefont {Sheehy}\ and\ \citenamefont
  {Schmalian}(2007)}]{sheehy07}%
  \BibitemOpen
  \bibfield  {author} {\bibinfo {author} {\bibfnamefont {D.}~\bibnamefont
  {Sheehy}}\ and\ \bibinfo {author} {\bibfnamefont {J.}~\bibnamefont
  {Schmalian}},\ }\href@noop {} {\bibfield  {journal} {\bibinfo  {journal}
  {Phys. Rev. Lett.}\ }\textbf {\bibinfo {volume} {99}} (\bibinfo {year}
  {2007})}\BibitemShut {NoStop}%
\bibitem [{\citenamefont {Sun}\ \emph {et~al.}(2010)\citenamefont {Sun},
  \citenamefont {Yan}, \citenamefont {Yao}, \citenamefont {Beitler},
  \citenamefont {Zhu},\ and\ \citenamefont {Tour}}]{sun10}%
  \BibitemOpen
  \bibfield  {author} {\bibinfo {author} {\bibfnamefont {Z.}~\bibnamefont
  {Sun}}, \bibinfo {author} {\bibfnamefont {Z.}~\bibnamefont {Yan}}, \bibinfo
  {author} {\bibfnamefont {J.}~\bibnamefont {Yao}}, \bibinfo {author}
  {\bibfnamefont {E.}~\bibnamefont {Beitler}}, \bibinfo {author} {\bibfnamefont
  {Y.}~\bibnamefont {Zhu}}, \ and\ \bibinfo {author} {\bibfnamefont {J.~M.}\
  \bibnamefont {Tour}},\ }\href {http://dx.doi.org/10.1038/nature09579}
  {\bibfield  {journal} {\bibinfo  {journal} {Nature}\ }\textbf {\bibinfo
  {volume} {468}},\ \bibinfo {pages} {549–552} (\bibinfo {year}
  {2010})}\BibitemShut {NoStop}%
\bibitem [{\citenamefont {Balan}\ \emph {et~al.}(2010)\citenamefont {Balan},
  \citenamefont {Kumar}, \citenamefont {Boukhicha}, \citenamefont {Beyssac},
  \citenamefont {Bouillard}, \citenamefont {Taverna}, \citenamefont {Sacks},
  \citenamefont {Marangolo}, \citenamefont {Lacaze}, \citenamefont {Gohler},
  \citenamefont {Escoffier}, \citenamefont {Poumirol},\ and\ \citenamefont
  {Shukla}}]{balan10}%
  \BibitemOpen
  \bibfield  {author} {\bibinfo {author} {\bibfnamefont {A.}~\bibnamefont
  {Balan}}, \bibinfo {author} {\bibfnamefont {R.}~\bibnamefont {Kumar}},
  \bibinfo {author} {\bibfnamefont {M.}~\bibnamefont {Boukhicha}}, \bibinfo
  {author} {\bibfnamefont {O.}~\bibnamefont {Beyssac}}, \bibinfo {author}
  {\bibfnamefont {J.-C.}\ \bibnamefont {Bouillard}}, \bibinfo {author}
  {\bibfnamefont {D.}~\bibnamefont {Taverna}}, \bibinfo {author} {\bibfnamefont
  {W.}~\bibnamefont {Sacks}}, \bibinfo {author} {\bibfnamefont
  {M.}~\bibnamefont {Marangolo}}, \bibinfo {author} {\bibfnamefont
  {E.}~\bibnamefont {Lacaze}}, \bibinfo {author} {\bibfnamefont
  {R.}~\bibnamefont {Gohler}}, \bibinfo {author} {\bibfnamefont
  {W.}~\bibnamefont {Escoffier}}, \bibinfo {author} {\bibfnamefont {J.-M.}\
  \bibnamefont {Poumirol}}, \ and\ \bibinfo {author} {\bibfnamefont
  {A.}~\bibnamefont {Shukla}},\ }\href
  {http://stacks.iop.org/0022-3727/43/i=37/a=374013} {\bibfield  {journal}
  {\bibinfo  {journal} {Journal of Physics D: Applied Physics}\ }\textbf
  {\bibinfo {volume} {43}},\ \bibinfo {pages} {374013} (\bibinfo {year}
  {2010})}\BibitemShut {NoStop}%
\bibitem [{\citenamefont {{Mendoza M.}}\ \emph {et~al.}(2013)\citenamefont
  {{Mendoza M.}}, \citenamefont {{Herrmann H. J.}},\ and\ \citenamefont {{Succi
  S.}}}]{mendoza13c}%
  \BibitemOpen
  \bibfield  {author} {\bibinfo {author} {\bibnamefont {{Mendoza M.}}},
  \bibinfo {author} {\bibnamefont {{Herrmann H. J.}}}, \ and\ \bibinfo {author}
  {\bibnamefont {{Succi S.}}},\ }\href {\doibase 10.1038/srep01052} {\bibfield
  {journal} {\bibinfo  {journal} {Sci. Rep.}\ }\textbf {\bibinfo {volume} {3}}
  (\bibinfo {year} {2013}),\ 10.1038/srep01052},\ \bibinfo {note}
  {10.1038/srep01052}\BibitemShut {NoStop}%
\bibitem [{\citenamefont {Bao}\ \emph {et~al.}(2009)\citenamefont {Bao},
  \citenamefont {Liu}, \citenamefont {Lei},\ and\ \citenamefont
  {Wang}}]{bao09}%
  \BibitemOpen
  \bibfield  {author} {\bibinfo {author} {\bibfnamefont {W.~S.}\ \bibnamefont
  {Bao}}, \bibinfo {author} {\bibfnamefont {S.~Y.}\ \bibnamefont {Liu}},
  \bibinfo {author} {\bibfnamefont {X.~L.}\ \bibnamefont {Lei}}, \ and\
  \bibinfo {author} {\bibfnamefont {C.~M.}\ \bibnamefont {Wang}},\ }\href@noop
  {} {\bibfield  {journal} {\bibinfo  {journal} {Journal of Physics: Condensed
  Matter}\ }\textbf {\bibinfo {volume} {21}} (\bibinfo {year}
  {2009})}\BibitemShut {NoStop}%
\bibitem [{\citenamefont {Anderson}\ and\ \citenamefont
  {Witting}(1974)}]{anderson74}%
  \BibitemOpen
  \bibfield  {author} {\bibinfo {author} {\bibfnamefont {J.}~\bibnamefont
  {Anderson}}\ and\ \bibinfo {author} {\bibfnamefont {H.}~\bibnamefont
  {Witting}},\ }\href {\doibase 10.1016/0031-8914(74)90355-3} {\bibfield
  {journal} {\bibinfo  {journal} {Physica}\ }\textbf {\bibinfo {volume} {74}},\
  \bibinfo {pages} {466–488} (\bibinfo {year} {1974})}\BibitemShut {NoStop}%
\bibitem [{\citenamefont {Mendoza}\ \emph {et~al.}(2013)\citenamefont
  {Mendoza}, \citenamefont {Karlin}, \citenamefont {Succi},\ and\ \citenamefont
  {Herrmann}}]{mendoza13a}%
  \BibitemOpen
  \bibfield  {author} {\bibinfo {author} {\bibfnamefont {M.}~\bibnamefont
  {Mendoza}}, \bibinfo {author} {\bibfnamefont {I.}~\bibnamefont {Karlin}},
  \bibinfo {author} {\bibfnamefont {S.}~\bibnamefont {Succi}}, \ and\ \bibinfo
  {author} {\bibfnamefont {H.}~\bibnamefont {Herrmann}},\ }\href@noop {}
  {\bibfield  {journal} {\bibinfo  {journal} {Journal of Statistical Mechanics:
  Theory and Experiment}\ } (\bibinfo {year} {2013})}\BibitemShut {NoStop}%
\bibitem [{\citenamefont {Bhatnagar}\ \emph {et~al.}(1954)\citenamefont
  {Bhatnagar}, \citenamefont {Gross},\ and\ \citenamefont {Krook}}]{bhat54}%
  \BibitemOpen
  \bibfield  {author} {\bibinfo {author} {\bibfnamefont {P.~L.}\ \bibnamefont
  {Bhatnagar}}, \bibinfo {author} {\bibfnamefont {E.~P.}\ \bibnamefont
  {Gross}}, \ and\ \bibinfo {author} {\bibfnamefont {M.}~\bibnamefont
  {Krook}},\ }\href {\doibase 10.1103/PhysRev.94.511} {\bibfield  {journal}
  {\bibinfo  {journal} {Phys. Rev.}\ }\textbf {\bibinfo {volume} {94}},\
  \bibinfo {pages} {511–525} (\bibinfo {year} {1954})}\BibitemShut {NoStop}%
\bibitem [{\citenamefont {Ansumali}\ and\ \citenamefont
  {Karlin}(2005)}]{karlin05a}%
  \BibitemOpen
  \bibfield  {author} {\bibinfo {author} {\bibfnamefont {S.}~\bibnamefont
  {Ansumali}}\ and\ \bibinfo {author} {\bibfnamefont {I.~V.}\ \bibnamefont
  {Karlin}},\ }\href {\doibase 10.1103/PhysRevLett.95.260605} {\bibfield
  {journal} {\bibinfo  {journal} {Phys. Rev. Lett.}\ }\textbf {\bibinfo
  {volume} {95}},\ \bibinfo {pages} {260605} (\bibinfo {year}
  {2005})}\BibitemShut {NoStop}%
\bibitem [{\citenamefont {Shan}(1997)}]{shan97}%
  \BibitemOpen
  \bibfield  {author} {\bibinfo {author} {\bibfnamefont {X.}~\bibnamefont
  {Shan}},\ }\href@noop {} {\bibfield  {journal} {\bibinfo  {journal} {Phys.
  Rev. E}\ }\textbf {\bibinfo {volume} {55}} (\bibinfo {year}
  {1997})}\BibitemShut {NoStop}%
\end{thebibliography}%

\end{document}